\pdfoutput=1
\documentclass[manuscript]{acmart}

\usepackage{todonotes}
\usepackage{rotating}
\usepackage{multirow}
\usepackage{natbib}
\usepackage{wasysym}
\usepackage{arydshln}
\usepackage{soul}
\RequirePackage[skins]{tcolorbox}
\newtcolorbox{custombox}[1]{
	colback=white,
	colframe=white,
	left=1mm,
	right=1mm,
	top=1mm,
	bottom=1mm,
	fonttitle=\bfseries,
	arc=0mm,
	leftrule=1mm,
	rightrule=0mm,
	toprule=0mm,
	bottomrule=0mm,
	notitle,
	before=\par\smallskip\noindent,
	before upper={\textbf{#1: } },
}

\AtBeginDocument{%
  \providecommand\BibTeX{{%
    \normalfont B\kern-0.5em{\scshape i\kern-0.25em b}\kern-0.8em\TeX}}}

\makeatletter
\newcommand\footnoteref[1]{\protected@xdef\@thefnmark{\ref{#1}}\@footnotemark}
\makeatother

\newcommand\AICodeGenerationTools{AI code generators}

\usepackage{menukeys}

\usepackage{caption}
\usepackage{subcaption}
\usepackage{multirow}

\copyrightyear{2024}
\acmYear{2024}
\setcopyright{rightsretained}
\acmConference[CHI]{CHI Conference on Computer-Human Interaction}
\acmBooktitle{CHI Conference on Computer-Human Interaction}
\acmDOI{11.1111/1111111.1111111}
\acmISBN{111-1-111-1111-1/11/11}

\begin{document}

\title[Interactions with Prompt Problems]{Interactions with Prompt Problems: A New Way to Teach Programming with Large Language Models}

\author{James Prather}
\orcid{0000-0003-2807-6042}
\affiliation{
  \institution{Abilene Christian University}
  \city{Abilene, Texas}
  \country{USA}
}
\email{james.prather@acu.edu}

\author{Paul Denny}
\orcid{0000-0002-5150-9806}
\affiliation{
  \institution{The University of Auckland}
  \city{Auckland}
  \country{New Zealand}
}
\email{paul@cs.auckland.ac.nz}

\author{Juho Leinonen}
\orcid{0000-0001-6829-9449}
\affiliation{
  \institution{The University of Auckland}
  \city{Auckland}
  \country{New Zealand}
}
\email{juho.leinonen@auckland.ac.nz}

\author{David H. Smith IV}
\email{dhsmith2@illinois.edu}
\orcid{0000-0002-6572-4347}
\affiliation{%
  \institution{University of Illinois at Urbana-Champaign}
  \city{Urbana, IL}
  \country{USA}
}

\author{Brent N. Reeves}
\email{brent.reeves@acu.edu}
\orcid{0000-0001-5781-1136}
\affiliation{%
  \institution{Abilene Christian University}
  \city{Abilene, Texas}
  \country{USA}
}

\author{Stephen	MacNeil}
\affiliation{
  \institution{Temple University}
  \city{Philadelphia}
  \state{PA}
  \country{United States}}
\email{stephen.macneil@temple.edu}
\orcid{0000-0003-2781-6619}

\author{Brett A. Becker}
\orcid{0000-0003-1446-647X}
\affiliation{
  \institution{University College Dublin}
  \city{Dublin}
  \country{Ireland}
}
\email{brett.becker@ucd.ie}

\author{Andrew Luxton-Reilly}
\orcid{0000-0001-8269-2909}
\affiliation{
  \institution{The University of Auckland}
  \city{Auckland}
  \country{New Zealand}
}
\email{a.luxton-reilly@auckland.ac.nz}

\author{Thezyrie Amarouche}
\orcid{0000-0003-3725-0049}
\affiliation{%
  \institution{University of Toronto Scarborough}
  \city{Toronto}
  \country{Canada}
}
\email{thezyrie.amarouche@mail.utoronto.ca}

\author{Bailey Kimmel}
\orcid{0009-0000-6655-0564}
\affiliation{
  \institution{Abilene Christian University}
  \city{Abilene, Texas}
  \country{United States}
}
\email{blk20c@acu.edu}

\renewcommand{\shortauthors}{Prather et al.}

\begin{abstract}

Large Language Models (LLMs) have upended decades of pedagogy in computing education. Students previously learned to code through \textit{writing} many small problems with less emphasis on code \textit{reading} and \textit{comprehension}. Recent research has shown that free code generation tools powered by LLMs can solve introductory programming problems presented in natural language with ease. In this paper, we propose a new way to teach programming with Prompt Problems. Students receive a problem visually, indicating how input should be transformed to output, and must translate that to a prompt for an LLM to decipher. The problem is considered correct when the code that is generated by the student prompt can pass all test cases. In this paper we present the design of this tool, discuss student interactions with it as they learn, and provide insights into this new class of programming problems as well as the design tools that integrate LLMs.

\end{abstract}


\begin{teaserfigure}
  \includegraphics[width=\textwidth]{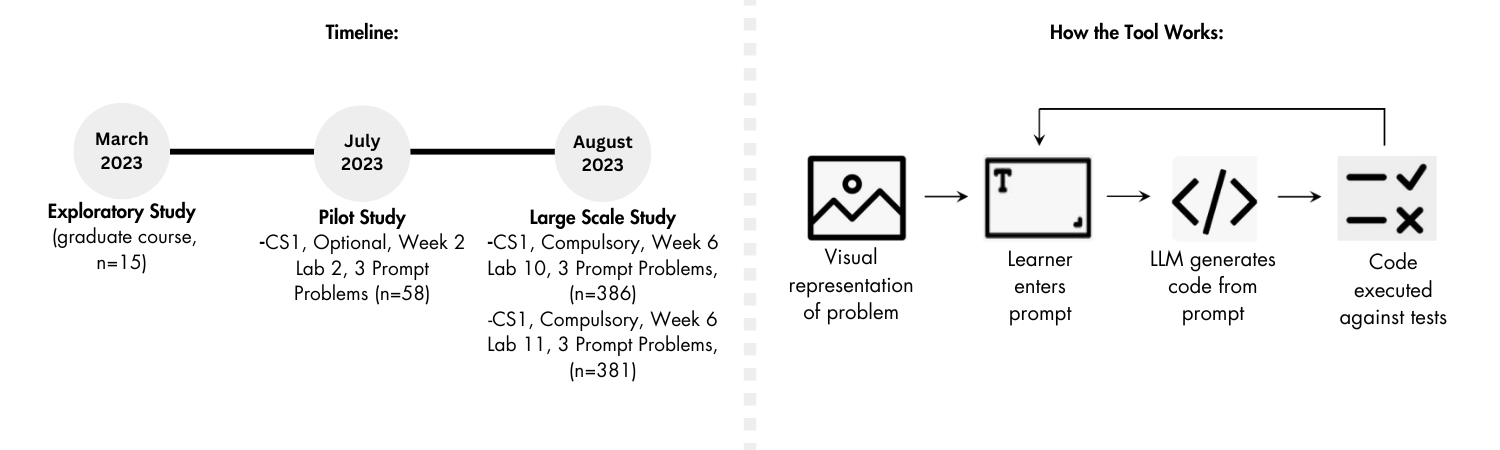}
  \caption{Left: An overview of the process of our data collection over three sequential studies. Right: How the Prompt Problems work in the tool we designed, \textsc{Promptly}.}
  \Description{An overview of the process of our data collection over three sequential studies (left) and how the tool we designed, \textsc{Promptly}, works (right). It shows three different studies over the course of 2023. \textsc{Promptly} works iteratively: students decompose a problem, write a prompt and submit it, review the code generated and the test cases that were run against it, and if not correct they revise their prompt accordingly.}
  \label{fig:teaser}
\end{teaserfigure}

\maketitle

%
%
\section{Introduction}
\label{sec:intro}

Generative AI models, such as the OpenAI GPT family, only entered to the mainstream in late 2022, but are already having a profound impact on how students learn programming. Researchers were quick to show that these models were surprisingly capable of solving most introductory programming assignments ~\cite{finnieansley2022robots,finnieansley2023my,savelka2023thrilled}. Now that benchmarks are being established \cite{prather2023WG-abstract} for their performance against typical computer science curricula, many are starting to turn their attention to the need for novel pedagogies and new types of assignments in the era of generative AI \cite{denny2023computing}. This is particularly pressing when introductory programming education has been focused on having students write dozens of small programs \cite{allen2019many, denny2011codewrite}, which now seems obsolete. Other learning activities with innovative interaction methods, such as Parsons problems \cite{ericson2022parsons, du2020review}, have been well studied in Human-Computer Interaction venues \cite{kelleher2019predicting, haynes2021parsonsproblems,weinman2021fadedparsons}. However, these are also at risk given the ability of AI models to solve them too \cite{reeves2023evaluating}. Educators have voiced concerns about academic integrity and that generative AI tools may cause students to become over-reliant on them ~\cite{becker2023programming,lau2023ban,zastudil2023generative}, which could lead to students having a diminished capacity to read and write code \cite{denny2023computing}. It is clear that skill sets required to learn programming are changing \cite{denny2023chat}, but it remains unclear how novices will develop them.

Many are predicting that programming as a discipline will shift away from \emph{writing} code itself toward generating code with prompt engineering, reading generated code, and then making small edits \cite{welsh2022end,bird2023taking}. However, early user studies have shown that while experienced programmers can effectively utilize generative AI tools \cite{vaithilingam2022expectation, barke2022grounded}, novice students often struggle with understanding code that they did not write but which the AI model generated for them \cite{prather2023tochi}. This underscores the need for new ways of teaching programming that scaffold the novice learning experience with generative AI without overwhelming them or threatening to undermine the process. Although this is becoming increasingly important, recent work has yet to describe an effective way to teach programming with prompts \cite{denny2023conversing}.

In this paper, we present ``Prompt Problems'' -- a novel way for students to interact with generative AI to learn programming. Students are presented with a visual, rather than textual, description of the problem and then must write a prompt that will generate the code to correctly solve it. After writing their prompt, a generative AI model creates the code, which we return to the student to see (but not edit). This code is then run against test cases for correctness. If their initial prompt did not return fully correct code, students must revise their prompt until the returned code can pass all test cases. In an ideal case, students are moving through multiple levels of abstraction, from problem decomposition to describing specifications, to code reading, to debugging, and back again. Over multiple studies (see Figure \ref{fig:teaser}), we recorded thousands of prompt submissions through our tool \textsc{Promptly}, which we examine here to better understand how students are interacting with this new class of programming problem. All studies were conducted following the ethical protocols for research of the university where data was collected\footnote{Ethics approval number [anonymized].}.


\subsection{Research Questions \& Contributions}
Our research questions are: 
\begin{enumerate}
    \item[\textbf{RQ1:}] How do students interact with Prompt Problems while learning programming?
    
    \item[\textbf{RQ2:}] How do students perceive Prompt Problems effecting their learning of programming concepts?
\end{enumerate}

There are three novel and important contributions of this work:
\begin{enumerate}
    \item We propose an innovative new type of programming problem to teach coding that utilizes large language models, which we call ``Prompt Problems''.
    
    \item We detail a tool we built to teach Prompt Problems, called ``Promptly'', discuss design considerations we made in the process, and what we learned about designing it that could help future generative AI tool creators.

    \item Through discussion of our exploratory study, pilot study, and large scale study, we detail student interactions with Prompt Problems and discuss how it might be used to teach programming in the era of generative AI.
    
\end{enumerate}

\textbf{Positionality statement:} We believe that LLMs are going to change computing education much like the calculator changed math education. Even though LLMs could suffer from certain drawbacks, such as over-reliance (as discussed below), we believe the benefits will outweigh them. LLMs are here to stay in our classrooms and will eventually come to be regarded as yet another tool for learning computing. This position informs our design and implementation of a new tool that can help students use LLMs while still learning to code.

%
%
\section{Related Work}
\label{sec:related-work}
In this section we review recent related work on large language models, their use in computing education, and prior user studies of \AICodeGenerationTools.

\subsection{Large Language Models}
Large language models (LLMs), especially those grounded on transformer architectures~\cite{vaswani2017attention}, have marked a significant advancement in the realm of natural language processing and artificial intelligence. A quintessential example of this is GPT-3, which represents a breakthrough in the field. The model was trained on an impressive 45 terabytes of textual data and boasts 175 billion model parameters~\cite{brown2020language}. The modus operandi during the training of LLMs is for the model to predict the next token given existing text. A token is essentially a word or a fragment of a word. The unparalleled abilities of these LLMs in tasks such as summarizing, translation, and answering questions are attributed mainly to the colossal amounts of training data and an increase in model size. Interaction with these models is typically done by submitting natural language prompts that instruct the model to generate some anticipated output. The output which is produced is typically non-deterministic meaning a single prompt can yield a variety of responses. This variability, sometimes termed ``creativity'', is a tunable parameter within the models that allows for control in the randomness of the model's responses.


In addition to natural language text, LLMs have been found to be proficient in generating source code. An early model specifically trained for this task is Codex~\cite{chen2021codex}, which is based on GPT-3 but fine-tuned with publicly available Python code found on GitHub. Codex also powers GitHub Copilot, which is one of the most widely used AI code generation tools. Fine-tuning the model on source code enhances the code generation capabilities of the model. Within the past year, due to the rapid progress in the capabilities of more generic models such as GPT-4, the Codex model has been deprecated, as generic text-to-text models such as GPT-4 surpass the capabilities of Codex in generating source code. In addition to closed source models such as Codex and GPT-4, open source models specifically aimed at code generation have been developed and released. These include, for example, StarCoder~\cite{li2023starcoder}, Code Llama~\cite{rozière2023code}, and CodeGen~\cite{nijkamp2023codegen}. The rapid emergence and adoption of these model further underscores the need for pedagogical approaches that teach students how to properly harness their capabilities.

\subsection{LLMs in Computing Education}
In recent years, large language models (LLMs) have garnered significant attention within computing education. This emergence of LLMs has resulted in numerous studies into their potential applications in computing education.

One fundamental area of interest has been the assessment of LLMs' performance in problem-solving tasks. Early work by Finnie-Ansley et al.~\cite{finnieansley2022robots} showed that Codex already outperformed the average student in introductory programming exercises that had been used in exams. A follow-up study found similar performance for more advanced programming exercises typically found in data structures and algorithms courses~\cite{finnieansley2023my}. The current state-of-the-art model, GPT-4, seems to excel in nearly all introductory programming tasks~\cite{savelka2023thrilled}. LLMs have also been shown to be capable of solving multiple-choice questions often found in introductory courses~\cite{savelka2023generative}. Now it has become clear that there is much more nuance in creating effective prompts than was believed.  A study  by Denny et al. found that by refining the prompts, the accuracy of GitHub Copilot in solving exercises could surge from approximately 50\% to about 80\%~\cite{denny2023conversing}. While this underscores the potential of LLMs, the study stops short of suggesting explicit methods to instruct students on effective prompting. Still, LLMs have their limitations. They have shown inconsistency in tasks requiring computational thinking~\cite{bellettini2023davinci} and reasoning about code~\cite{savelka2023large}, and their performance in solving Parsons problems (where code blocks need to be arranged to the correct order) trails behind their performance in solving traditional programming problems~\cite{reeves2023evaluating}. Furthermore, their grasp on more advanced concepts remains to be seen.

Beyond problem-solving, the potential of LLMs to support both learners and educators has been a focal point of research. An early exploration by Sarsa et al. revealed that Codex can generate unique personalized programming exercises and help explain code in natural language, proving beneficial for novice programmers~\cite{sarsa2022automatic}. Subsequent studies, such as those by MacNeil et al. and Leinonen et al., validated that code explanations rendered by LLMs are not only appreciated by students~\cite{macneil2023experiences} but are also, on average, superior to those crafted by the students themselves~\cite{leinonen2023comparing}.

Further extending their utility, LLMs have been studied for their capability in formulating learning objectives for courses~\cite{sridhar2023harnessing}. Moreover, LLMs show promising performance in responding to students' help requests~\cite{hellas2023exploring}, which could be especially beneficial for large courses such as MOOCs, where teaching assistant resources might be scant. There also exists some exploratory work that has delved into the use of LLMs for providing feedback~\cite{pankiewicz2023large,kiesler2023exploring}. LLMs have been found to be a useful tool for enhancing programming error messages~\cite{leinonen2023using}, which are notoriously complex for novice programmers to understand~\cite{becker2019compiler}.

However, while the benefits LLMs bring to computing education are promising, there are inherent challenges to integrating LLMs into computing education. Concerns about students becoming overly dependent on LLMs~\cite{prather2023tochi,bommasani2022opportunities} or misusing them in prohibited scenarios~\cite{becker2023programming}, essentially committing academic misconduct, have been raised. 

Altogether, LLMs are reshaping the landscape of computing education~\cite{denny2023chat,denny2023computing}.  As the field continues to evolve, the careful integration of LLMs into computing education, while balancing their benefits with potential risks, remains paramount.

\subsection{User Studies of LLMs and LLM-Based Tools}
Most early work on LLMs in computing education has focused on expert evaluation, where LLM outputs are evaluated by the researchers without having users interact with models. However, in the past year, more and more research has evaluated LLMs and LLM-based tools with students. One subarea where user studies have been conducted is studying how programmers use LLMs to generate code. In a study by Vaithilingam et al.~\cite{vaithilingam2022expectation}, programmers were given three programming problems in Python that they had to solve in Visual Studio Code. In two of the tasks, only the built-in IntelliSense auto-completion functionality was used by the participants, while in one problem they were given access to GitHub Copilot. The majority of the participants preferred using Copilot over the built-in auto-complete feature, mentioning time-saving as a positive benefit. However, there were no actual differences in the time spent programming between using Copilot and using IntelliSense. Barke et al. found that the use of LLM code generation could be split into two modes, acceleration and exploration~\cite{barke2022grounded}. In acceleration, users had specific goals where they used the LLM to help them achieve those goals, whereas in exploration, the participants used LLMs to help them explore different ways to go about solving the problem.

Recently, more work has emerged that has focused solely on novice programmers. Jayagopal et al.~\cite{jayagopal2022exploring} found that being able to user LLMs to support programming was exciting to novice programmers, and that students would engage in prompt engineering to try generate code if the original code did not match their expectations. Kazemitabaar et al.~\cite{kazemitabaar2023studying} conducted a between-subject study where half of the students could use Codex while programming while the other half could not. They found that the group that could use Codex performed better, and solved tasks faster. In addition, having used Codex did not hurt that group's performance in subsequent manual code modification tasks, even when Codex was not available. Prather et al.~\cite{prather2023tochi} observed novice programmers using Copilot for the first time. They found that in addition to the exploration and acceleration interaction patterns identified by Barke et al.~\cite{barke2022grounded}, novices also engaged in less productive interaction patterns they dubbed `shepherding' and `drifting'. In shepherding, novices spent the majority of time on trying to prompt the model to generate the code, rarely modifying the created code manually, which could be a sign of over-reliance on LLM support. In drifting, students would drift from one (incorrect) Copilot suggestion to the next, not committing to any suggestion. This might be a sign of possibly not understanding or trusting the suggestions generated by Copilot.

\subsection{Explain in Plain English}
Since Prompt Problems require students to explain the problem well enough that an LLM can generate correct code, it stands adjacent to the long line of research on ``Explain in Plain English'' (EiPE). In 2006, Whalley et al. found a relationship between programming ability and the ability of students to explain code in simple terms in their own words \cite{whalley2006australasian}. From there, EiPE research continued to be validated over the course of the next decade \cite{lopez2008relationships, venables2009closer, lister2009further}. Whether about introductory programming or more advanced topics, these studies all point to the fact that EiPE utilizes a unique approach to learning programming that may reach students who otherwise struggle \cite{murphy2012ability}. In 2014, Corney et al. found that EiPE problems "are likely to improve student ability to reason about code and, by extension, improve student ability to write code" \cite{corney2014explain}. In 2023, Cruz et al. found that students still need more practice in reading and comprehending code, such as through EiPE exercises \cite{cruz2023exploring}. Recent work by Li et al. has shown that AI systems can be used to provide feedback on EiPE programming problems, but that false positives (where a tool wrongly says a student answer is correct) can have a detrimental effect on learning \cite{li2023wrong}. This study by Li et al. paves the way for the next logical step in assignments that utilize EiPE. We avoid the limitation of false positives identified in Li et al. with Prompt Problems by running the generated code against a suite of test cases to ensure correctness.

\section{Designing a Tool for Assessing Prompt Generation}
\label{sec:designing-the-tool}

\subsection{Exploratory Study}
We conducted an exploratory study to analyze the types of challenges novice programmers might face when they use large language models to generate source code for programming tasks. We recruited 36 computer science graduate students at [institution anonymized] to participate in the study in March 2023. The participants were given visual representations of a coding problem, without any text that they could copy and paste into their prompt, and they were tasked to create a prompt to ChatGPT that would create the code to solve the problem. Their experiences were documented through asking for their reflections of the activity, and they had the option to share their ChatGPT logs, which 15 students did.

The analysis of these 15 student interactions revealed three major trends: (1) initially, many students provided vague prompts without enough detail to actually solve the problem; (2) upon facing issues, students would often try to clarify their intent through additional messages to the LLM rather than restructuring their original prompts; (3) and when faced with errors from the code generated by ChatGPT, they directly asked the LLM to fix the errors rather than trying to diagnose the cause themselves. Despite their advanced programming background, many students struggled with effective prompting. In the written reflections, students commented that it was difficult to instruct ChatGPT to generate the correct code, which provided further indication that explicit instruction and practice at prompting at the introductory level will be required in order to help students use LLMs for code generation successfully. This adds the skill of ``prompting'' to the pantheon of programming skills, with the difference that for other skills, assessment items already exist. In an effort to provide a method of allowing students to practice this skill and allowing instructors to assess it we introduce ``Prompt Problems'' delivered via our tool, \textsc{Promptly}, to address this gap.

\subsection{The Tool: \textsc{Promptly}}

\subsubsection{Implementation}
Our concrete implementation of the tool uses React and NodeJS as its key frameworks, and Material design for the styling of UI components. The client-side React implementation is accessible via Firebase Hosting, and the Express (NodeJS) backend is powered by Firebase Functions, operating within a serverless framework.  The backend communicates with OpenAI's API and transmits responses to a  JobeInABox\footnote{\href{https://github.com/trampgeek/jobeinabox}{github.com/trampgeek/jobeinabox}} sandbox which is hosted on an EC2 AWS instance.  We explored the use of several specific OpenAI models, including \emph{text-davinci-003} and \emph{gpt-3.5-turbo}. Our current implementation uses \emph{text-davinci-003} which, although now officially a legacy model, is less likely to generate superfluous text and comments in the responses.  We found that the \emph{gpt-3.5-turbo} model requires significant additional prompting to increase the likelihood of generating only executable code, but that relying on prompting alone can be unreliable.  Future work will explore additional filtering approaches in order to transition to a newer model. All relevant data, including prompts, responses, and testing outcomes are stored using Firestore's NoSQL database. Students must login to the system using their university IDs, but no private student data is shared with GPT.

\begin{figure}
\centering
  \includegraphics[scale=0.56]{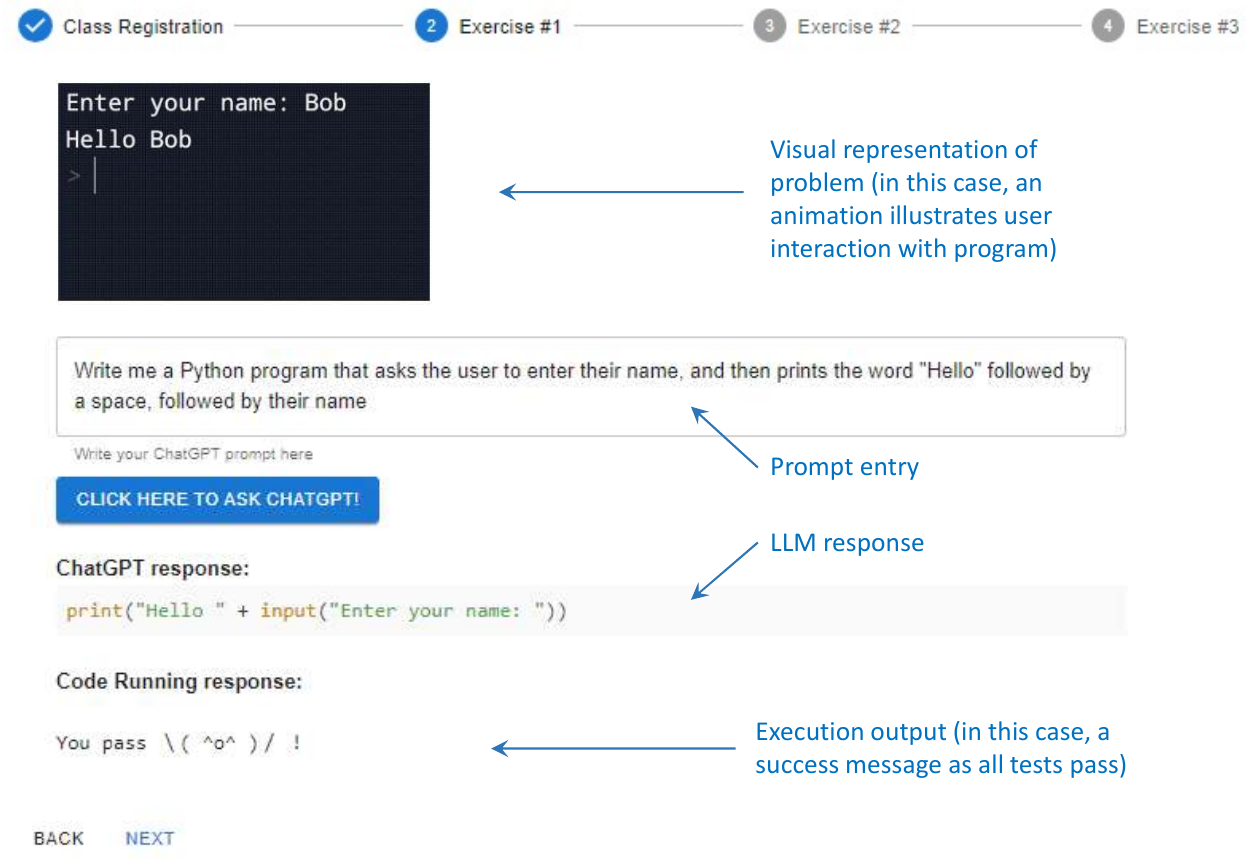}
  \caption{The tool we created, \textsc{Promptly}, showing the user solving the problem by submitting a prompt.}
  \label{fig:tool-screenshot}
\end{figure}

\subsubsection{Organization of problems}
Within the \textsc{Promptly} tool, sets of Prompt Problems are organized into course repositories, which students can select after logging in.  Each Prompt Problem within a course repository consists of a visual representation of a problem -- that is, an image that does not include a textual description of the problem -- and a set of associated test cases that are used to verify the code that is generated by the LLM.  

Prompt Problems for a given course are presented in order, and a student can navigate through these using `Back' and `Next' buttons.  Once a Prompt Problem is selected, the student is shown the visual representation of the problem, and a partial prompt to complete.  For problems where the solution is a Python program, this partial prompt begins: ``Write a Python program that...'', which provides guidance to the student.  If the problem requires students to write a single function, then the partial prompt is: ``Write a Python function called...''.  As soon as any text for extending the prompt is entered by the student, the ``Click here to ask ChatGPT!'' button is enabled.  Clicking this button constructs a prompt that is sent to the LLM. This prompt consists of the verbatim text entered by the student, as well as some additional prompting to guide the model to produce only code and no additional explanatory text.  

Once the code response is received from the LLM, it is then sent to a sandbox for execution against the set of test cases.  We use the publicly available sandbox associated with the CodeRunner tool \cite{lobb2016Coderunner}.  If the generated code passes all of the test cases for the prompt problem, then the student receives a success message and is directed to progress to the next problem.  If any of the test cases fail, then the first failing test case is shown to the student.  
This design decision encourages students to focus on resolving one issue at a time, mirroring advice frequently given to students around fixing syntax errors \cite{becker2018fix}, and serves to prevent students from feeling discouraged if their prompt generates code that fails a large number of tests.
At this point, they are able to edit the prompt and resubmit in order to generate a new code response. 

Figure \ref{fig:tool-screenshot} shows a screenshot of the tool interface.  The following instructional message is shown but not included in the screenshot: \emph{``Your task is to view the visual representation of the problem and then type a prompt which describes the task sufficiently well for the language model to generate a correct solution in Python. If the code that is generated is not correct, you will see test output below the coding area and you can try again by modifying the prompt!''}. In the screenshot shown in Figure \ref{fig:tool-screenshot}, the learner has just submitted a successful prompt. In this case the student can then continue to submit prompts to the same problem or click the "NEXT" button, which is now highlighted in blue. Students must solve each problem in order.

\subsubsection{Design considerations}
After the exploratory study, we discussed design approaches to the tool, opting for simplicity and iteration above all else. The only student interaction is typing and submitting their prompt. Since the goal of the tool is to have students learn to successfully prompt for the correct code by sufficiently describing the problem, we decided they should not be able to edit the generated code nor should they edit the test cases. This design would ideally force students to decompose the problem, read the code generated, compare it with the test cases to discern why it is failing, and then update their prompt accordingly. However, there are other possible ways to design Prompt Problems, which we discuss below in Section \ref{sec:discussion}.



\begin{figure}
\centering
\begin{minipage}{.5\textwidth}
  \centering
  \includegraphics[width=.7\linewidth]{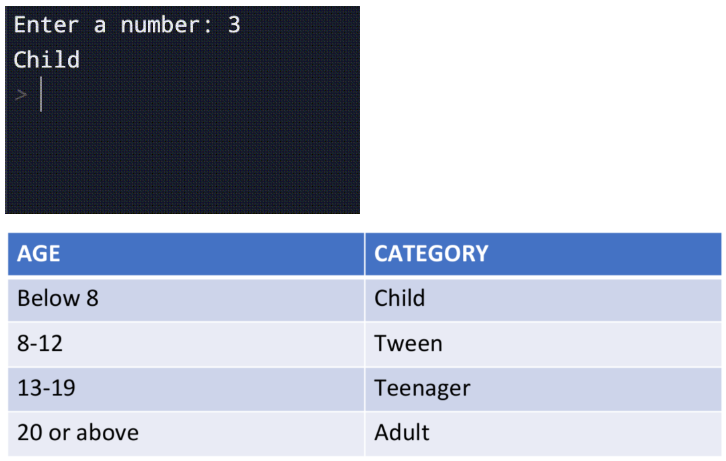}
  \captionof{figure}{Producing a categorization based on age.}
  \label{fig:ex2_question}
\end{minipage}%
\begin{minipage}{.5\textwidth}
  \centering
  \includegraphics[width=1.0\linewidth]{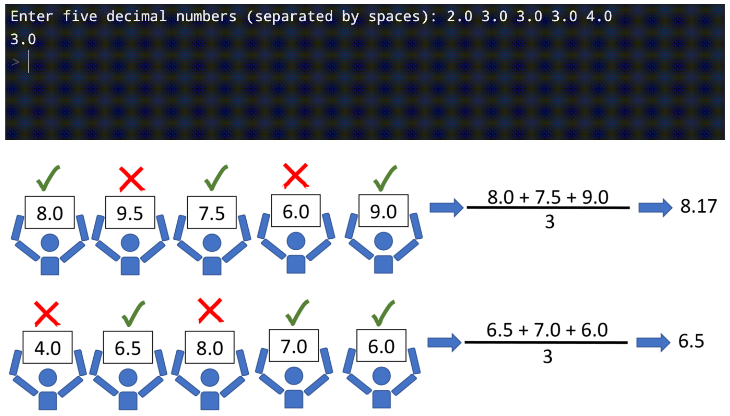}
  \captionof{figure}{Calculating the average of the ``middle'' values out of a set of five values (using the metaphor of judges scoring an athletic competition, where the highest and lowest values are excluded).}
  \label{fig:ex3_question}
\end{minipage}
\end{figure}

%
%

\section{Pilot Study}
\label{sec:pilot-study}
We ran a pilot study of \textsc{Promptly} in July 2023, in a first-year programming course (CS1) taught in Python at \textit{[institution anonymized]} in \textit{[country anonymized]}. Students in this course typically have no (or limited) prior programming experience. 
Participation was optional and 58 (14\%) of 414 students enrolled chose to attempt at least one question using \textsc{Promptly}. 




Prompt Problems are a novel type of problem for students and we were interested in understanding student interactions when using \textsc{Promptly} to solve problems, and their perceptions of the tool. Prompt Problems are also new to instructors, so we wanted to learn  how to effectively deliver Prompt Problems.
We organized our pilot around the following two goals that match the research questions of this work: 

\begin{description}
\item Goal 1: How do students interact with \textsc{Promptly} in terms of the lengths of their prompts and their success rates?
\item Goal 2: What are students' perceptions of \textsc{Promptly}, and the process of constructing prompts for LLMs to solve programming problems? 
\end{description}


One of the problems asked students to develop a program that would ask the user for their name as input and then print ``Hello'' followed by the name input (see Figure \ref{fig:tool-screenshot}). The second was to classify an input age using one of four labels (Figure \ref{fig:ex2_question}), and the third was to output the average of the three middle values (after removing the min and max) of five floats provided as input (Figure \ref{fig:ex3_question}). For each problem, a visual representation included a short ($\sim$10 second) animation shown as a command-prompt style window. The animation illustrated example user input followed by the desired output, and we found that showing keystrokes being entered by the user was helpful for differentiating input from output.  Designing a prompt problem is similar in some respects to designing a traditional programming exercise -- we start by identifying the general problem to be solved and then craft a model code solution.  From this model solution, we develop a test suite (a set of input/output pairs) that can be used to verify generated code, much in the same way that a test suite would be designed for a programming task in a typical automated assessment tool.  Developing the visual representation is in some cases simple (e.g. in Figures \ref{fig:tool-screenshot} and \ref{fig:ex2_question} one or more inputs and corresponding outputs appear in the image), and in other cases requires designing an image that conveys the problem to be solved (e.g. the judges score cards in Figure \ref{fig:ex3_question}).  We provide a more thorough discussion of the considerations for designing prompt problems in Section \ref{sec:designingpromptproblems}. 

In terms of interaction, for each problem we recorded the number of students that were ultimately successful, the mean number of prompt submissions required to solve the problem and the mean number of words in the prompts. In terms of student perception, students could provide feedback on their experience with \textsc{Promptly} -- students were asked: \emph{``We would appreciate hearing about your experiences completing the exercises and in particular, how you think the experience of writing prompts may help you to learn programming''}. 


\subsection{Student Interactions}
\label{sec:pilot_interactions}
\begin{figure}[!h]
\centering
  \includegraphics[width=\linewidth]{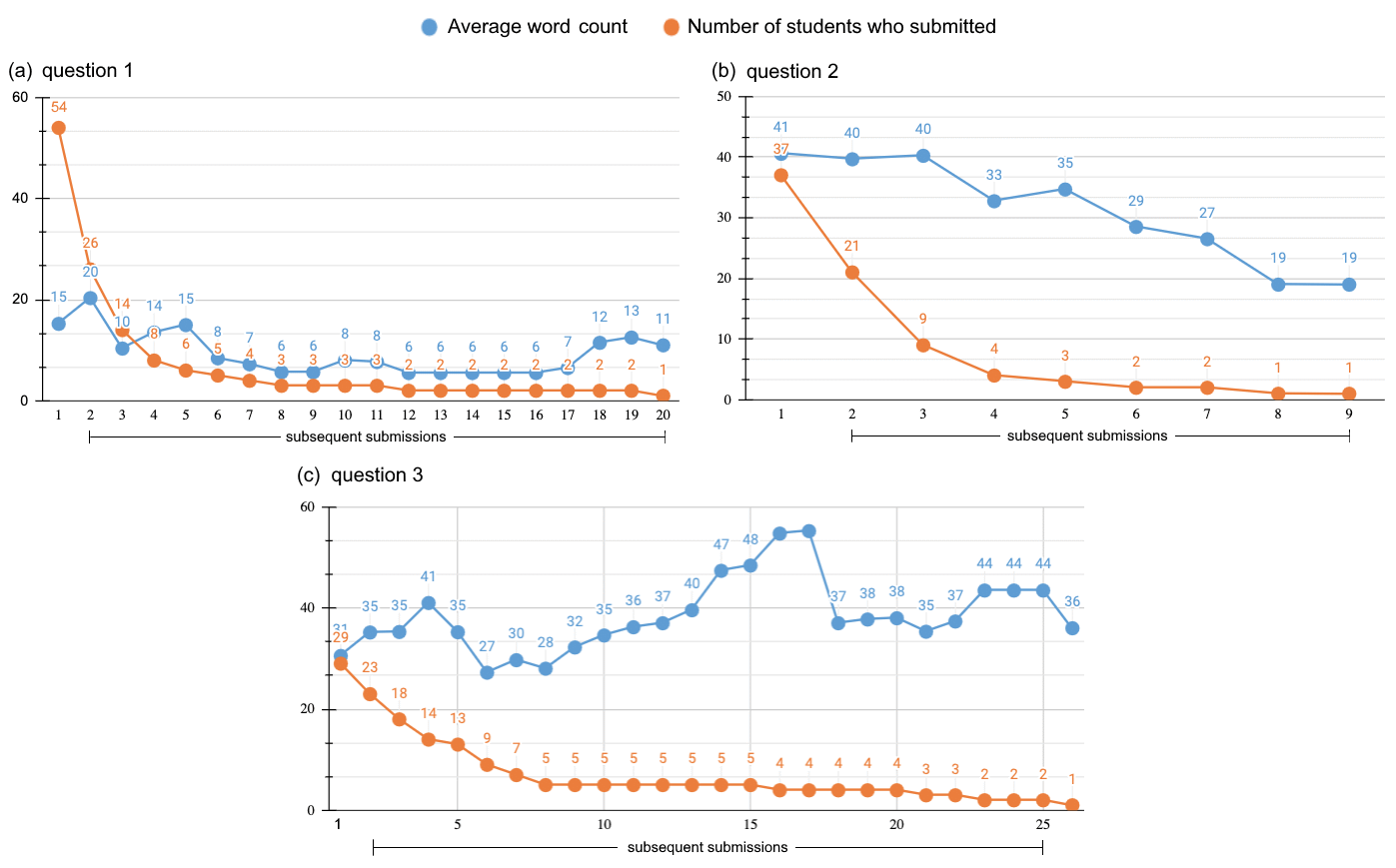}
  \caption{The average number of words in each subsequent submission and number of participants that submitted. On the x-axis, 1 is the initial submission (attempt) per question and 2- are subsequent submissions (attempts).}
  \label{fig:q1-3_avgwords-vs-numStudentSub}
\end{figure}

On average participants made the following number of attempts for problems 1-3 respectively: 2.7, 2.16, and 6.4. Initially this could lead to the conclusion that problem 3 caused the most difficulty. Figure \ref{fig:q1-3_avgwords-vs-numStudentSub} (a)-(c) shows for problems 1-3 respectively, how the mean word count of prompts and the number of students submitting them, change through subsequent submissions. As an example Figure \ref{fig:q1-3_avgwords-vs-numStudentSub} (a) shows that 54 students made an initial submission (1 on the x-axis), and that the average word count of these prompts was 15. As students find success (or give up) on the problem, fewer students make subsequent submissions. Comparing problems 1 and 2, mean prompt length generally decreases with subsequent submissions with the exception of the final small-n `bump' at submissions 18-20 for problem 1. For problem 3, the mean prompt length tends to increase slightly over subsequent submissions, but like with problem 1 there is an anomaly towards the small-n tail (submissions 18-26). 

\subsection{Student Perceptions}
\label{sec:pilot_reflections}

We analyzed responses from the 58 students who answered the reflection question. We report the main themes that emerged from our analysis below:

\begin{itemize}
  \item \textbf{Exposure to new coding constructs:} Given that our evaluation was conducted early in the course, the code that was generated would sometimes contain features that were unfamiliar to students.  For the most part, students commented positively on this aspect, and a theme emerged around the way the tool introduced students to new programming constructs and techniques.  As one student commented: \emph{``These exercises introduced me to new functions... so this method of writing code could help increase my programming vocabulary''}. 
  
  \item \textbf{Enhancement of computational thinking:} We also found students valued the way in which the tool challenged them to think carefully about how to solve a problem and communicate precise specifications: \emph{``You would have to understand what the question is asking to be able to give a working prompt to the AI, so it seems very good for making you think properly about the question''} and \emph{``Writing the prompts can help you with visualizing the steps required in the programming''}.
  
  \item \textbf{General positive feedback:} 49\% of the participants expressed positive sentiments about \textsc{Promptly}, and this was the most common theme overall. Students wrote: \emph{``I think that is was a good for practicing asking AI''} and \emph{``Asking AI to write prompts help a lot in programming!!!''}. 

  \item \textbf{General negative sentiment:} 11\% of the participants expressed negative sentiments towards generative AI in general, saying they valued creativity, that they refuse to use tools like ChatGPT, or expressing fear they might become over-reliant on generative AI tools. These fears appear in the literature discussed above~\cite{barke2022grounded}.
\end{itemize}

Overall, while most students reported finding \textsc{Promptly} beneficial, particularly for exposure to new programming constructs and for strengthening computational thinking skills when communicating a problem, a minority of students were hesitant about the use of generative AI tools for learning programming.

\subsection{Lessons Learned}
There were three lessons from the pilot study that informed our implementation of the large scale study detailed below in Section \ref{sec:large-scale-study}. 

First, most students in the pilot study appreciated prompt problems which could expose them to new ideas, programming constructs, and help them with program decomposition and specification writing. This helped us verify that Prompt Problems delivered via \textsc{Promptly} were useful for programming education and that we should proceed.

Second, some students seemed to try prompts without much thought, just to see what happens. This behavior is similar to what we have observed with automated assessment tools \cite{becker2019compiler}, thus we did not see this as a deterrent to using Prompt Problems.  Instead, we attempted to design new problems with an eye toward minimizing this behavior by having the problem presentation include a useful keyword as the function call. Hopefully, this gives students a starting place, rather than beginning by throwing things against the wall to see what sticks.

Third, we learned that writing effective Prompt Problems was more difficult than expected due to a balance that needs to be struck. Problems should be understandable to students, however given the goals we have for \textsc{Promptly} it is not desired for it to be too easy to construct a prompt which results in a correct solution. At the same time, we also do not want students to spend too much time deciphering the visual presentation of the problem nor do we want them to get hung-up on the challenges of prompting. We designed \textsc{Promptly} to help students as a scaffold, leading them to the correct answer in successive steps. All of the problems we showed the CS1 students involved writing prompts for entire programs, which proved more difficult than we had expected, both for students and in terms of implementation. We therefore believe that problems involving functions struck the right balance by imposing a structure that allowed the prompt process itself to be an effective scaffolding `bridge' between the problem and an eventual correct answer. We decided to bring this lesson forward to a larger-scale study.

%
%
\section{Large-Scale Study}
\label{sec:large-scale-study}
The large-scale study was conducted in the same CS1 course as the pilot study described earlier in Section \ref{sec:pilot-study} but approximately three weeks later in August 2023. 

\subsection{Methods}
\begin{figure}
     \centering
     \includegraphics[width=\linewidth]{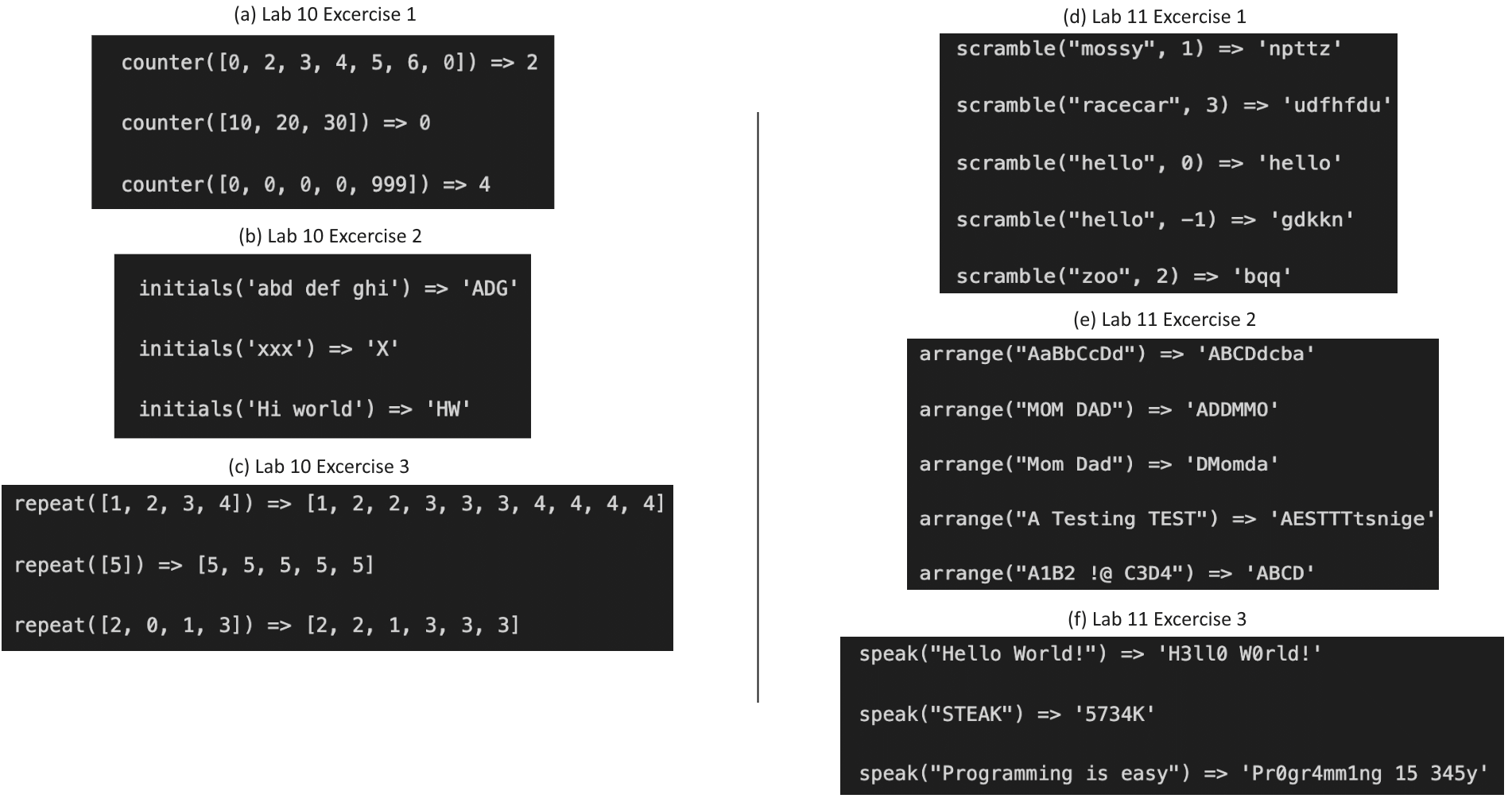}
     \caption{Lab 10 Exercises 1 (a), 2 (b), 3 (c) and Lab 11 Exercises 1 (d), 2 (e), 3(f)}
     \label{fig:lab10_11}
\end{figure}

Exercises using \textsc{Promptly} were introduced as a component of laboratory 10, and included in laboratory 11.  At this point in the course, students had already learned about variables, operators, expressions, conditions, loops, sequences, and functions.  For each laboratory, students were asked to complete a set of three exercises using \textsc{Promptly}, and then complete a short reflective exercise.  The questions in laboratory 10 were designed to introduce students to the process of prompt generation using relatively straightforward problems.  In laboratory 11, students were provided with more challenging questions that were designed to require more complex prompts.  The questions were designed to have code solutions that are similar in content and complexity to the code-writing tasks in these labs. These questions for both labs 10 and 11 are shown in Figure \ref{fig:lab10_11}.  Unlike the pilot study, students were not shown additional graphics.  The reflective exercise used in both laboratory 10 and laboratory 11 asked students to respond to the questions \textit{``Consider how easy or difficult it was to specify the problem that you wanted to solve, if there were any special cases that needed more attention, and whether you understood the code that was produced.  (A) Do you think this tool should continue to be used in programming courses?  (B) What impact do you think it would have on your learning?''}

In total, 386 students attempted lab 10, and 381 students attempted lab 11. Each lab contributes 1\% towards the final grade, and the \textsc{Promptly} exercises comprised approximately 10\% of the marks available for each lab, totaling approximately 0.2\% of the final grade.  Given this, not all students attempted them. Only 202 (52\%) students in lab 10 and 147 (39\%) in lab 11 attempted the \textsc{Promptly} exercises. Given the relatively low impact these problems had on course grades these percentages are somewhat expected. However, because some students in the pilot study indicated resistance to ever using AI and we did not want to miss their insights, we still analyzed all open response questions, which we evaluate in our qualitative results below.

We examine the resulting data using a mixed-methods approach. For quantitative results, we present basic statistical data on student interactions with the tool for each question in both labs. We also use a data-driven approach to examine interactions in more detail below in Section \ref{sec:results:quant}.

%
%
%

For qualitative data, three of the authors analyzed the reflective comments submitted by students using reflexive thematic analysis~\cite{BraunClarke2022TA, Braun2006ReflexTA}, which aims to understand patterns and meaning present in qualitative data.  This uses an approach that is exploratory and develops codes fluidly through an iterative process of revising codes as researchers become more familiar with the data~\cite{braun2019reflexive}.  Because codes develop throughout the process, inter-coder reliability measures are not typically calculated for reflexive thematic analysis, but instead reliability of the results is achieved through other means.  To ensure reliability in our analysis, we held several group meetings where authors compared codes and discussed differences in the application of codes, and combining and splitting codes as our understanding of the data increased and the themes began to emerge, as is appropriate during reflexive thematic analysis~\cite{mcdonald2019irr}. We chose reflexive thematic analysis for our qualitative data because it was recently used by one of the first papers to examine novice programmer interactions with LLMs at an HCI venue \cite{prather2023tochi}.  We conducted the analysis in several phases, as outlined by Braun and Clark~\cite{Braun2006ReflexTA}:
\begin{enumerate}
\item Familiarization with the data: We began by reading through a sample of the responses and made notes about initial observations and potential codes. 
\item Generate codes: Subsequently, the researchers involved in the analysis met to discuss their initial codes and examples that exemplified the codes.  The codes were split and combined, applied to a larger subset of data, and then discussed further.  The final set of codes was then applied to 50\% of the overall data obtained from students, at which point saturation was reached and the researchers observed no new insights from further coding.
\item Development of themes: The codes were grouped into themes and these themes were discussed and refined. Examples of data that exemplified the potential themes were identified.
\end{enumerate}

\begin{table*}[htb]
\small
\caption{Summary of student interactions with the Prompt Problems in Labs 10 and 11.  For students, we provide the total number of unique students that attempted each problem (Total), the number who got it correct (Correct), and the number who got it correct on the first try (First Try). For submissions, we provide the total number of prompt submissions made for that problem (Count), the mean number of submissions (Mean), the minimum number of submissions any student had to correctly solve the problem (Min), and the maximum number of submissions any student had whether correct or incorrect (Max). To describe the words in submitted correct prompts, we provide the average number of words in correct prompts (Mean), the minimum number of words in correct prompts (Min), and the maximum number of words in correct prompts (Max).}


\begin{tabular}{c|ccc|cccc|ccc}
\label{tab:large-quant-summary}
\textbf{Problem} & \multicolumn{3}{c|}{\textbf{Students}} & \multicolumn{4}{c|}{\textbf{Submissions}} & \multicolumn{3}{c}{\textbf{Words in Prompts}} \\
 \phantom{x} & Total & Correct & First Try &  Count &  Mean &  Min &  Max &  Mean &  Min &  Max\\
\toprule
Lab10-1 & 202 & 118 & 32 & 884 & 4.37 & 1 & 30 & 25.79 & 7  & 76  \\
Lab10-2 & 108 & 108 & 74 & 212 & 1.96 & 1 & 10 & 27.39 & 8  & 93  \\
Lab10-3 & 107 & 104 & 67 & 224 & 2.09 & 1 & 20 & 34.78 & 8  & 119 \\
\midrule
Lab11-1 & 147 & 105 & 39 & 491 & 3.34 & 1 & 27 & 41.11 & 9  & 198 \\
Lab11-2 & 97  & 82  & 20 & 502 & 5.17 & 1 & 28 & 43.96 & 16  & 86  \\
Lab11-3 & 80  & 60  & 5  & 626 & 7.82 & 1 & 47 & 54.07 & 23 & 115 \\ \bottomrule
\end{tabular}
\end{table*}

%
%
\subsection{Results - Quantitative}
\label{sec:results:quant}
In order to answer our first research question, \emph{``RQ1: How do students interact with Prompt Problems while learning programming?''}, we turn to our quantitative results.

There were 3,077 prompt submissions using \textsc{Promptly} for all six questions across both labs. After cleaning the data and removing prompts that were copy/paste dumps from other contexts or malformed attempts, 2,939 prompt submissions remained. Table \ref{tab:large-quant-summary} shows the descriptive statistics for students submitting, their submissions, and word counts in those submissions. Although the premise of \textsc{Promptly} was to have students write English language prompts to solve each problem, many students wrote code instead. Some students used a mixture of English natural language and code.  We found it helpful to categorize prompts based on the approach used by the student. Three researchers examined several hundred prompts and discussed categorization before landing on the three used: English, expression, code. One researcher then tagged every submission with one of the three labels. A submission was considered ``English'' if it contained no code. It was tagged as ``expression'' if it had an English sentence and code, such as function definitions. Therefore, an expression could be a long paragraph of natural language sentences with only one function definition or it could be a mostly complete function to handle the problem with a sentence asking the AI to create something. A prompt was tagged as ``code'' if it contained only code with no natural language sentences.

\begin{figure}
     \centering
     \includegraphics[width=\linewidth]{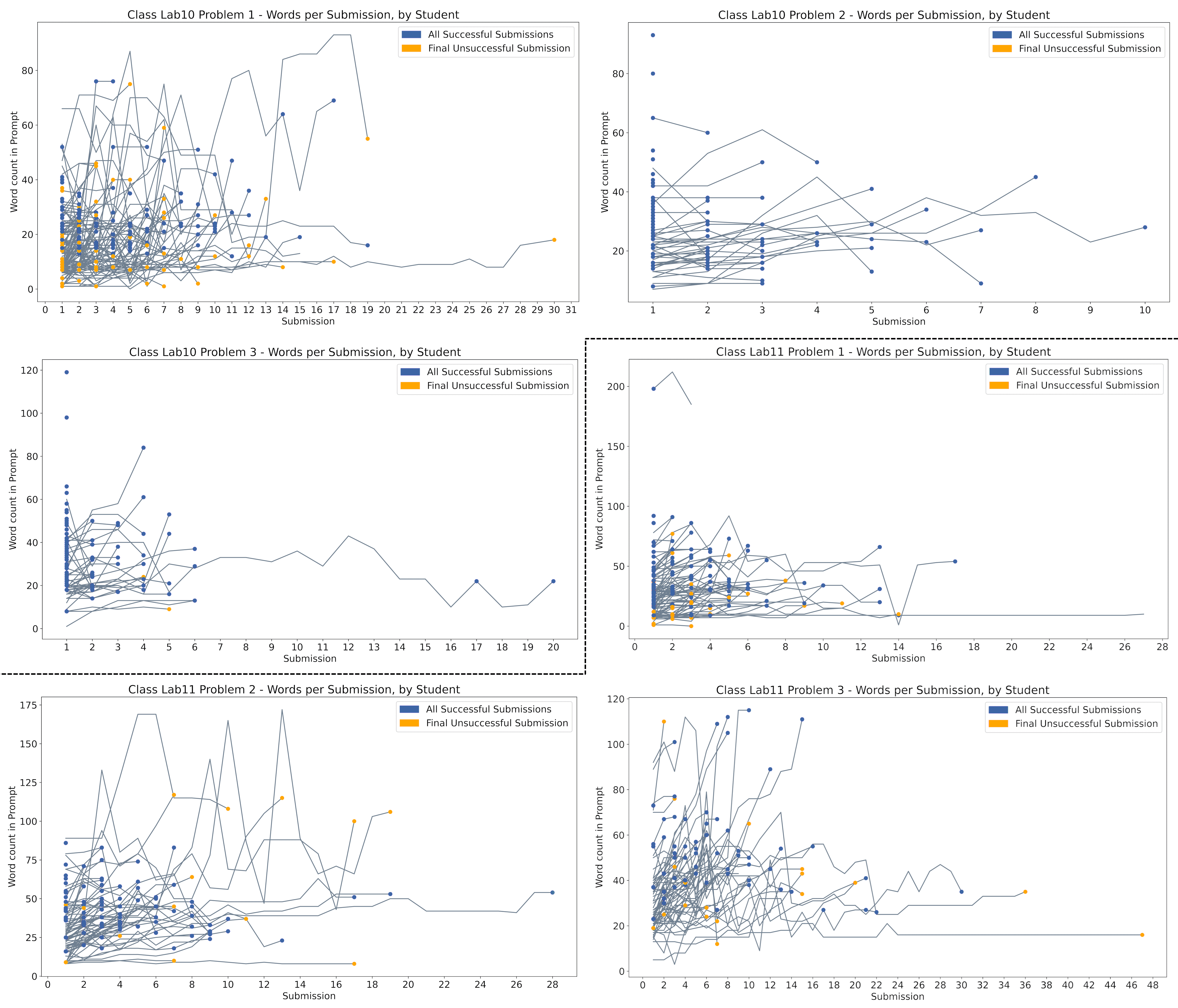}
     \caption{Submission streams for all six questions across both labs. Each line represents all submissions made by a student for that problem. Green dots denote every successful submission; red dots denote final unsuccessful submission.}
     \label{fig:all-labs-submission-streams}
\end{figure}

Some students were successful with extremely terse submissions while others wrote excessively long submissions. One example of this is lab 10, question 2 (see Table \ref{tab:large-quant-summary} and the top right of Figure \ref{fig:all-labs-submission-streams}). This problem was about inputting a string and extracting the first character of each word (delimited by spaces) and then outputting that as concatenated capital letters. The shortest and longest successful prompts were:

\begin{description}
    \item[P175] ``Write me a Python function called initials(words)''
    \item[P298] ``Write me a Python function called initials which accept the string 's' as its sole input. To create a new string, the function needs to initialise the first letter of each word in the input string, convert it to uppercase, and then concatenate these uppercase initials. This new string should be returned as the result. The function should return 'ADG' if the input text is 'abd def ghi'. The function should return 'X' if the supplied string is 'xxx'. The function must return 'HW' if the given string is 'Hi world'.''
\end{description}

\begin{figure}
     \centering
     \includegraphics[width=\linewidth]{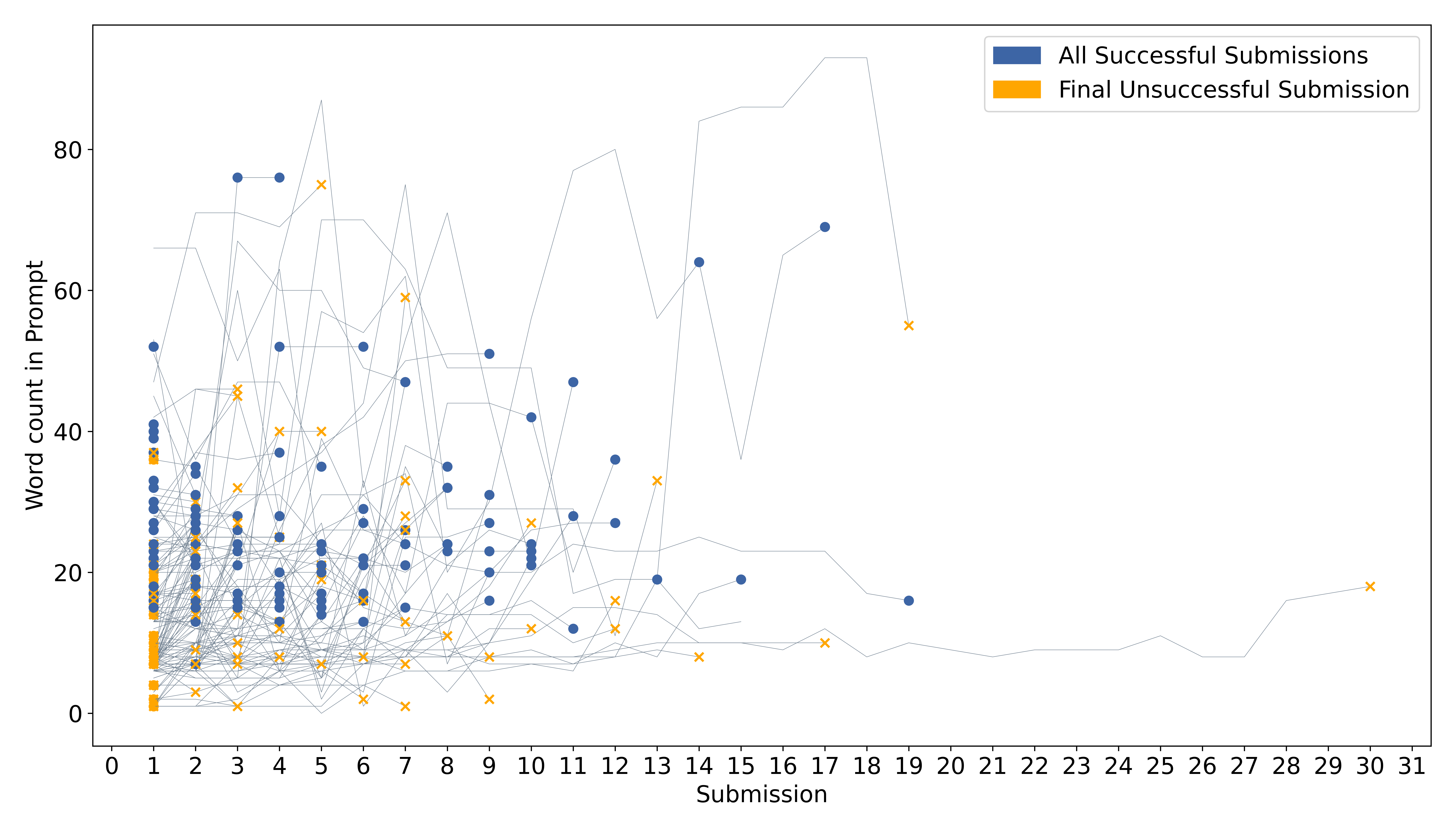}
     \caption{Submission streams for lab 10 problem 1. Each line represents all submissions made by a student for that problem. Green dots denote every successful submission; red dots denote final unsuccessful submission. This is the same graph as shown in the top left of Figure \ref{fig:all-labs-submission-streams}, only enlarged.}
     \label{fig:lab10-problem1}
\end{figure}

Figure \ref{fig:lab10-problem1} shows the student submission streams for Lab 10 Problem 1, i.e. the number of submissions that each student made by word count over time where green dots denote a correct submission and a red dot denotes a final incomplete submission. Figure \ref{fig:all-labs-submission-streams} shows the same data for all six problems. These graphs provide a window into the interesting journeys that students took while trying to solve each problem. After a burst of activity at the beginning, most students either made a correct submission or gave up. However, some students persevered. Although most students who submitted many more prompts than the mean did not successfully solve the problem, there were some students who did. 
Below we examine student interactions with \textsc{Promptly} for those students who submitted more times than two standard deviations above the mean. Several students submitted more than one successful prompt to the same problem, indicating experimentation with the problem and tool. We also examine some of their interactions with \textsc{Promptly} below to get a better sense of what they were doing.

\subsubsection{The long tail}
\label{subsec:quant-long-tail}
Students who made more submissions than two standard deviations above the mean represent students who generally refused to give up until either getting it correct or perhaps becoming exasperated by the process. The average number of submissions across all six problems was 4.3 with a standard deviation of 5.6. Due to space constraints, we chose to examine this across all problems in both labs, rather than by individual problem. We consider some interesting cases below.


Participant 240, who submitted 30 prompts to problem 1 (as seen on the far right of Figure \ref{fig:lab10-problem1}), started with an English language prompt and tweaked it twice more before moving to expressions. These expressions started asking for functions with specific names. Over the next 23 submissions, the student tried many variations of function names in an apparent attempt to coax the model into creating the correct function. For instance, four subsequent submissions in this part of the stream were ``Write me a Python function called get\_number\_of\_zeroes():'', then ``Write me a Python function called 0\_in\_lists\_counter():'', then ``Write me a Python function called find\_zero\_character():'' and finally ``Write me a Python function called counter\_the\_zero\_characters\_in\_lists():''. The last four prompts this student submitted were in English again, indicating the student taking a new approach. Their final prompt was ``Write me a Python function which counts the number of 0s in a list of integers called counter''. The final attempt was still unsuccessful and the student gave up, which did not contain enough detail to pass all test cases.

Participant 125 submitted 19 times to problem 1 and was eventually successful (the lower of the two dots with 19 attempts in Figure \ref{fig:lab10-problem1}). They started with an expression, briefly attempted English for one prompt, and then returned to expressions again up to submission 17, which was ``Write me a Python function calling counter([0,2,3,4,5,6,0]) >= 2 that returns the number of zeros by themselves''. This was merely the latest variation on the same prompt and was not working. The student tried an expression once more on submission 18 before pivoting to English with the prompt ``Write me a Python function calling counter that returns the number of zeros by themselves'' which was successful.

From the very beginning, Participant 74 submitted code prompts to problem 3, making a total of 19 submissions, two of which were correct (as seen on Figure \ref{fig:all-labs-submission-streams}, middle left). This student eventually succeeded on submission 16 by writing the entire function they wanted the model to generate. The code in the very next submission was significantly stripped down with only the function name and parameters, as if the student attempted to see if they could get the model to generate the correct solution from less code. When this did not work, their next submission included the next line of code under the function definition. This still did not work, so the student resorted to resubmitting their correct (code) prompt a final time.

Many of the other examples include students who attempted English until switching to expressions by concatenating test case failure feedback onto their English language prompts. Contrary to what we expected, a common strategy for these students with long submission streams was to try variations on ideas by changing a few words, rather than continually adding more words. This can be seen in Figure \ref{fig:all-labs-submission-streams} on particularly long submission streams that, other than some outliers, remain around the same word length the for entire stream.

\subsubsection{Playing with \textsc{Promptly}}
\label{subsec:quant-playing}
Many students played with \textsc{Promptly} by submitting additional prompts even after they had correctly solved the problem. A few interesting cases are considered below.

Participant 40 submitted 12 times to lab 10 problem 1. Starting with ``Write me a Python function called poo'' followed by the prompt ``what is this'', this student seemed to be playing with the tool. The next seven prompts were more serious attempts where the student slowly built their prompt using more and more test case data. Submission 10 continued this trend and was successful: ``create a function that takes a list and counts the number of times zero occurs
here is an example of inputs: counter ([0, 2, 3, 4, 5, 6, 0]) => 2, counter([10, 20, 30]) => 0, counter([0, 0, 0, 0, 999]) => 4''. However, the student did not stop there and next attempted an English prompt: ``write a python function that readds [sic] a list of numbers, and counts the number of 0, and returns this number''. When this was not successful, they added the test case data back into their new English prompt, which returned a correct solution.

Participant 171 submitted 26 times to lab 11 problem 1. Interestingly, this student got a correct prompt on submission 13. The first 13 submissions were all variations on the same expression prompt, with only the word inside the quotes in the function call changing. For instance, submission 12 was, ``Write me a Python function called scramble("mossy", 1)''. Submission 13, which was correct, continued this trend and was ``Write me a Python function called scramble("hello", 0)''. It seems that the student benefited from the randomness inherent in probability models. This appears to have confused or intrigued the student because the next 13 prompts were all like the previous ones with only small variations on the word inside quotes in the function call. However, none of these were successful. It appears the student was attempting to replicate their good luck because many of these subsequent prompts were the exact same as their successful one, but variance in the model did not generate the same code again.

Participant 243 submitted 22 times to lab 11 problem 3 (as seen in Figure \ref{fig:all-labs-submission-streams}, bottom right). This student had three successful submissions. The first 14 submissions submissions were variations on the same sentence, such as this one from submission 14: ``Write me a Python function called speak that replaces all letters with numbers that look visually similar''. At this point, the student seemed to recognize edge cases their prompt was not addressing and changed their prompt to include replacing `T' with '7'. Finally, on submission 17 they landed on their first successful prompt: ``Write me a Python function called speak. The function will replace all letters with numbers that look similar and T or t will be replaced with 7''. They then tried to make a shorter prompt: ``Write me a Python function called speak. The function will replace all letters with numbers that look identical'', which was not successful. They then brought back the more specific part of the successful prompt: ``Write me a Python function called speak. The function will replace all letters with numbers that look identical. T or t will be replaced with 7'', but this was not successful either. The student then returned to their successful prompt. Finally, they made one last submission, attempting to reduce the number of words by removing ``all'', which was also successful.

Another prevailing pattern in this interaction type was that students would often attempt to reduce the total number of words in their prompt. The average delta in word length from the first successful prompt to the next prompt for the same problem was -3.15. 
Students who spent less than 1 hour on problems averaged just over 6 minutes with a mode of 3.2 minutes.  Ten students spent over an hour on some of the problems, and twelve students spent over 2 hours on some of the problems.

%
%
\subsection{Results - Qualitative}
\label{sec:qual-results}
\begin{figure}
    \centering
    \includegraphics[width=\linewidth]{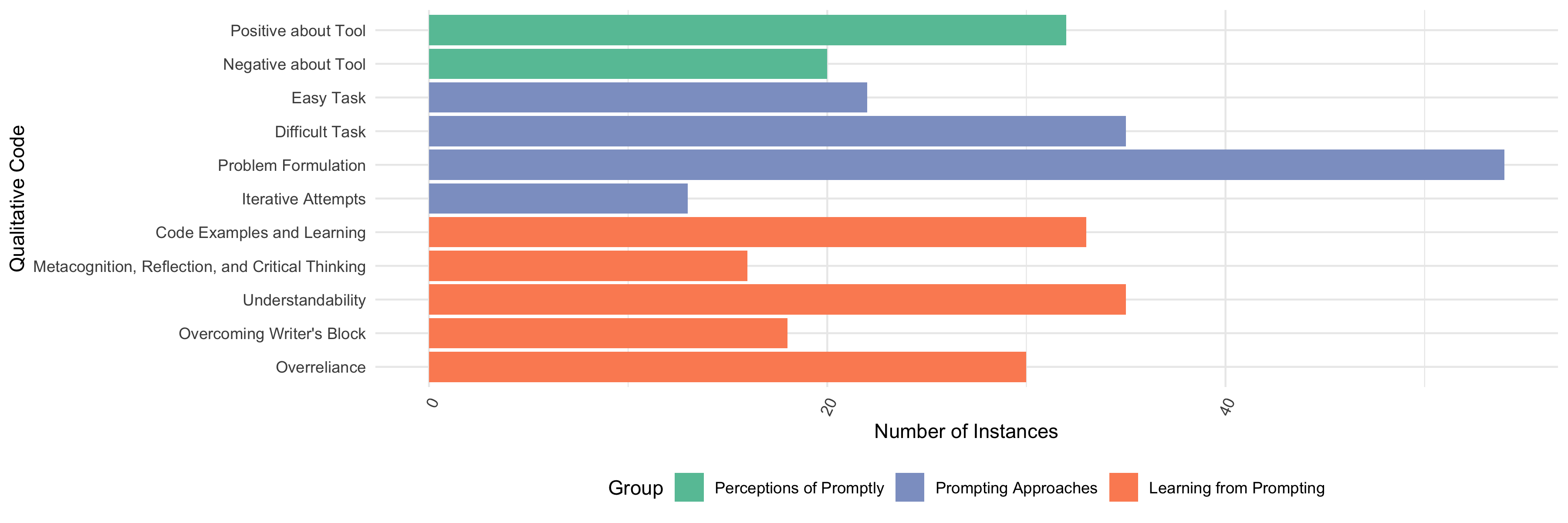}
    \caption{An overview of the relative occurrences of codes applied during the qualitative analysis.} 
    \label{fig:qualitative-summary}
\end{figure}

In order to answer our second research question \emph{``RQ2: How do students perceive Prompt Problems effecting their learning of programming concepts?''}, we turn to the qualitative analysis of the student responses to the reflective questions. Table \ref{tab:codebook} summarizes the three main themes,  corresponding codes, and definitions of the codes. The relative occurrence of the qualitative observations are summarized in Figure~\ref{fig:qualitative-summary}.

\begin{table}[]
\begin{tabular}{lll}
\hline
Theme                                                                               & Code                                                                       & Code Definition                                                                                                                                                                                                       \\ \hline
\multirow{2}{*}{\begin{tabular}[c]{@{}l@{}}Perceptions of \\ the Tool\end{tabular}} & \begin{tabular}[c]{@{}l@{}}Positive Sentiment \\ Towards Tool\end{tabular} & \begin{tabular}[c]{@{}l@{}}Student expresses a form of positive sentiment\\ towards Promptly as a tool (e.g., UI, feedback).\end{tabular}                                                                             \\ \cline{2-3}
                                                                                    & \begin{tabular}[c]{@{}l@{}}Negative Sentiment \\ Towards Tool\end{tabular} & \begin{tabular}[c]{@{}l@{}}Student expresses a negative sentiment towards\\ Promptly as a tool.\end{tabular}                                                                                                          \\ \hline
\multirow{4}{*}{Prompting Approaches}                                                          & Easy Task(s)                                                               & \begin{tabular}[c]{@{}l@{}}Student indicates that the task of successfully\\ generating a prompt was easy.\end{tabular}                                                                                               \\ \cline{2-3}
                                                                                    & Difficult Task(s)                                                          & \begin{tabular}[c]{@{}l@{}}Student indicates that the task of successfully\\ generating a prompt was difficult.\end{tabular}                                                                                          \\ \cline{2-3}
                                                                                    & Problem Formulation                                                        & \begin{tabular}[c]{@{}l@{}}Students response discusses their approach or \\ experiences with formulating a description of \\ the problem.\end{tabular}                                                                \\ \cline{2-3}
                                                                                    & Iterative Attempts                                                         & \begin{tabular}[c]{@{}l@{}}Student discusses their experience with or approach\\ to iterative modifications to their initially incorrect \\ prompt(s) or creation of successive new prompts.\end{tabular}             \\ \hline
\multirow{5}{*}{\begin{tabular}[c]{@{}l@{}}Learning from \\ Prompting\end{tabular}} & Code Examples and Learning                                                 & \begin{tabular}[c]{@{}l@{}}Student mentions learning from seeing different\\ approaches to the solution, as generated via ChatGPT.\end{tabular}                                                                       \\ \cline{2-3}
                                                                                    & Metacognition                                                              & \begin{tabular}[c]{@{}l@{}}Student mentions their thought processes.\end{tabular}                                                                        \\ \cline{2-3}
                                                                                    & Understandability                                                          & \begin{tabular}[c]{@{}l@{}}Student mentions aspects of the generated code \\ that improves or limits their ability to interpret it.\end{tabular}                                                                      \\ \cline{2-3}
                                                                                    & Overcoming Writers Block                                                   & \begin{tabular}[c]{@{}l@{}}Students indicate that prompting, as a skill, is or could\\ be useful for situations where they know the task to\\ perform but are unsure of how to form an initial solution.\end{tabular} \\ \cline{2-3}
                                                                                    & Overreliance                                                               & Student indicates they might come to rely on prompting.                                                                                                                                                               \\ \hline
\end{tabular}
    \caption{The table of themes, codes, and code definitions.}
    \label{tab:codebook}
\end{table}

%
%

\subsubsection{Perceptions of \textsc{Promptly}}

The first theme captured student responses to the actual tool, including affective responses and observations about the user interface or interactions. In evaluating students' perceptions of the tool, we coded responses as having a positive or negative sentiment towards the tool and captured any feedback about the user interface or interactions.  We identified relatively few comments about the tool, most of which were rather generic.

Comments involving positive sentiment tended to express a general preference to use the tool in the future.

\begin{quote}
    ``I have used the generative AI tool to complete the three exercises and found the experience to be quite interesting and valuable. It allowed me to see how prompts can be used to guide the AI in generating code for specific tasks based on input and output examples. The tool was relatively easy to use, and I appreciated the ability to experiment with different prompts to achieve the desired code output.'' (P183)
\end{quote}
    

A small number of comments suggested new features that could be added, or identified restrictions on the way they interacted with the tool that decreased usability. 

\begin{quote}
    ``The tool is helpful in some cases, but unlike actually using ChatGPT, the tool does not take previous prompts into consideration, which makes adjusting the code more difficult than it should be.''  (P209)
\end{quote}



\subsubsection{Prompting approaches}
\label{subsec:qual-prompting}

We observed participants often commented on their experiences with prompt generation task, including perceptions of difficulty and strategies for formulating and refining prompts. Responses that included a description of how easy or difficult students found the task of creating a successful prompt were coded as easy or difficult, respectively. The most common observation from our qualitative analysis was instances where students described how they formulated prompts and the need to be specific in their phrasing of the prompts. In many of these instances, students also talked about the difficulty of translating from a problem brief into a prompt that would be understood by the LLM. 


Students were mixed about the difficulty of the prompt engineering task with slightly more students characterizing it as being difficult than easy. Students who found it easy tended to describe the need to be specific and clearly define the requirements. 

\begin{quote}
        ``I think specifying the problems I wanted to solve was quite easy, as all you have to do is to concisely write down what you want the code to do, and it provides a really good answer.'' (P45)
\end{quote}

When describing the challenges they faced prompting the LLM, students talked about the difficulty in specifying exactly what they wanted. This included being able to describe the problem, as stated by P365, ``It was hard to put into words what the program would actually do.'' It also included being able translate that problem into a prompt: 

\begin{quote}
    ``For Question 3, I was unable to solve it. I understood the problem, but describing it via prompts was more difficult.'' (P319)
\end{quote}

Students faced difficulties finding the right level of specificity: 

\begin{quote}
    ``... I found it difficult to specify the problem that I wanted to solve, I understand that it needs to be specific, however, the response I would get from ChatGPT did not help me understand the code at all.'' (P57)
\end{quote}

These difficulties led to significant frustration for students. Students described how even when they believed that they had correctly specified the prompt, there were still instances when the model would misinterpret the meaning: 

\begin{quote}
    ``Finding a perfect answer which suits the requirements is very tedious, I have to learn how to prompt-engineer and understand it further in detail, but when the prompt is crafted the way it wants, sometimes it misreads it which causes frustration.'' (P52)
\end{quote}

A big part of this difficulty appeared to stem from a lack of appropriate vocabulary that could appropriately guide the model: 

\begin{quote}
    ``I found it quite difficult as I did not have the vocabulary to express what I wanted ChatGPT to do as I did not know the names of some of the inbuilt functions and stuff in python.'' (P61)
\end{quote}

Students also grappled with the non-deterministic nature of generative AI tools describing how sometimes they would get the right answer and other times they would need to iterate on their prompts. Some students described this experience as being frustrating and others saw it as part of the process of working with generative AI tools:  

\begin{quote}
    ``It was medium difficulty to specify the problem that I wanted to solve. There were special cases where the prompt can give me the correct answer but generating a new code with the same prompt can give me an incorrect answer. Therefore, I had to go into more detail to solve the problem.'' (P161)
\end{quote}
\begin{quote}
    ``Despite AI writing the code for me, I still found myself thinking logically about what I wanted the function to do and how I wanted it to do it, in order to articulate this to the AI.'' (P299)  
\end{quote}

Some students described this iterative process as similar to the debugging work that they do when working in lab and writing code. P227 described how this iteration was needed when they received a response that used arithmetic operations to shift ASCII values within the ranges of uppercase and lowercase letters. 

\begin{quote}
    ``Writing these prompts felt similar to creating code for these labs---I would present an initial solution, see what doesn't work and why, and fix it. One of the prompts produced an answer which I didn't understand - it used numbers to manipulate letters.'' (P227)
\end{quote}

While some students made the analogy between iterative prompting and debugging code, students did not necessarily enjoy this debugging experience. And in one case highlighting that this iterative interaction ended up not being worth the effort when compared with just writing the code themselves: 

\begin{quote}
``...I spent a lot of time pedantically arguing with the AI about case sensitivity in one of the questions: (I tried every variant I could think of, like: `... and answer is not case sensitive' `...which understands text regardless of case and returns accordingly' and so on and so forth, and in the end when I continued not to get what I needed (the AI would only do one case, or only do for vowels, or do some strange translation of every character) then I specified the exact characters I wanted changed in both upper and then lowercase.(`....where aeiost is 43057 respectively, and AEIOST is 43057 respectively...') This frustrated me because it was effectively like writing the code myself.'' (P240) 
\end{quote} 

In this particular instance, P240 highlights the specificity required in order to obtain the desired results. To further understand the challenges that P240 described related to case sensitivity, we viewed their activities logs. We confirmed that they tried two strategies: 1) articulating the task for the model explicitly, or 2) providing examples of input-output pairs, akin to few-shot learning. When the descriptions failed, P240 provided the example: ``Write me a Python function called speak which converts text regardless of case as follows: a to 4, e to 3, t to 7, i to 1, s to 5.'' This did not work. P240 reverted to their previous strategy of describing the program's intended functionality, which also did not work. Finally, P240 provided examples again specifying to convert ``text aeiost to 431057 respectively and AEIOST to 431057 respectively.'' Using this final approach, the test cases passed.

Finally, participants described how the challenges were also about needing to interpret the responses and identify edge cases or `loopholes' that would produce the wrong answer even when the code looks correct: 

\begin{quote}
    ``You must specify exactly what your code can accept, or the AI may leave loopholes that will produce errors or incorrect results.'' (P385)
\end{quote}

In this case, P385 specified numerous detailed rules the model should follow; however, the rules contained faulty logic. One rule stated the characters should ``Keep all uppercase letters at the beginning of the string in their original order'' but they should have been alphabetically sorted. They later gave correct test cases which had the uppercase letters sorted. These few shot examples conflicted with the previously stated rule. However, we observed across multiple activity logs where students missed a test case and were able to adapt their prompts to account for that case, often appending a description of the corner case. For example P22 specified ``and t looks like 7'' when the test case failed. In these cases, monitoring the responses from the model often required not only identifying edge cases but also understanding how the resulting code works and making connections between the prompt and the resulting code:

\begin{quote}
    ``Manipulating a prompt to generate a correct program took several attempts to rephrase and understand the specifics of what the code required.'' (P22)
\end{quote}

These results indicate that students faced difficulty crafting prompts that the LLM could comprehend, primarily due to struggles in achieving the appropriate level of specificity and a deficiency in their domain specific vocabulary. Students who were successful described being able to clearly articulate the instructions to the model and eagerly engaging in an iterative debugging process where they re-articulated their prompts until they got the correct response. Finally, students needed to closely monitor the responses from the models to account for the non-deterministic nature of the models and the potential for missing edge cases.

\subsubsection{Learning from generated code}
\label{subsec:qual-learning}

The third and final theme that emerged during the data analysis revolves around learning, and the perceived impact of the tool. Students reported that they learned by seeing examples of different solutions to a problem, reflecting on the problems themselves and different ways they can be described, use of terminology and precision, comprehensibility of code and how that impacts learning, the value of prompting for progressing in programming tasks and the potentially negative impact on learning from focusing on prompts rather than code.

Perhaps the most common learning benefits listed by students was the ability to see different approaches to solving a given problem.
\begin{quote}
    ``I think it is a way of gaining `knowledge' on how a code can be written and the logic processes that a person can use when writing their own code. Somewhat like learning from the writing style of your tutor and they ended up with the code they had written. I think A.I and chatgpt is a good way of expanding this kind of coding `creativity' for a lack of a better word. Gaining experience in the different ways that a person can write code.'' (P257)
\end{quote}

Students, such as P109, described this ability to ``see a different view point on solving the problem'' as a benefit. P297 described a new feature of the tool that could facilitate comparisons, ``Maybe this tool could be used after a question is completed so that you can compare your solution to ChatGPT's.'' This new perspective also challenged students to reflect on the generated code and underlying concepts: 

\begin{quote}
    ``During this activity, I noticed that the generated code can be quite complex, especially for more intricate problems. It's important to review and analyze the generated code to grasp the underlying concepts.'' (P385)
\end{quote}


Similarly, many students noted that becoming proficient in prompting could be used in cases where they are stuck on forming a solution to a problem or overcoming an initial ``writers block'' while coding.
\begin{quote}
    ``I think AI learning tools are great for compsci courses as they can streamline things like bugfixing and code making - you no longer need another person to tell you what is going wrong, and it is much easier to get out of `writers block' for coding'' (P135)
\end{quote}

Interestingly, students also reported that solving prompt problems raised their awareness of their own cognitive processes by thinking about problems and the solutions, or reflecting on their own code. For example, P233 described how it required them to focus on problem formulation while working toward their solution: 

\begin{quote}
    ``This tool has a notable impact on my learning, particularly in fostering critical thinking. It prompts me to think deeply about problem formulation and effective ways to communicate with the tool to get desired outcomes.'' (233)
\end{quote}

Another student described how it helped them to think about programming differently, focusing on understanding the problem and planning a solution before writing it, ultimately making them more aware of what they were doing: 

\begin{quote}
    ``I think it has a positive impact on my learning because it gives me a different way to think about programming and gets me thinking about what I actually want the code to do before I start writing it. I think this will improve my coding as it will make me more aware of what I need to do. It is also a good test to check that I can identify patterns in code and give instructions based on this knowledge, which I think requires a different level of understanding.'' (241) 
\end{quote}



These two examples show how students described engaging in metacognitive aspects of learning such as planning their problem solving approach and monitoring whether they understood what they were doing. This increased awareness was also exemplified by students who described how the tool might better support reflecting on their learning, for example P381 said, 


\begin{quote}
    ``It made me think a lot about what the function was actually doing, and I think it can be useful to practice communicating, explaining, and comprehending code. I believe it has a positive effect on my personal learning when used effectively as a tool of reflection.'' (P381) 
\end{quote}

This sentiment was echoed by many students and it appears that the automated nature of these tools did not detract from that experience of reflection. 

\begin{quote}
    ``Despite AI writing the code for me, I still found myself thinking logically about what I wanted the function to do and how I wanted it to do it.'' (P299)
\end{quote}

However, despite these perceived benefits, several students mentioned potential caveats. For example, many stated that the tool often generated code that uses functions that they were unfamiliar with. This limited their ability to interpret what the generated code was doing and learn from the activity.

\begin{quote}
    ``Having tried more exercises than last lab with different prompts I am aware of some flaws with using AI tools that I previously overlooked. For example Chat GPT used functions that I have not used before like: ord() chr() list() filter() sorted() as the foundation for the codes process. As a result I did not learn much from the exercise.'' (P105)
\end{quote}

While P105 described not learning from advanced concepts, there were some students who viewed this as an opportunity to expand their knowledge and engage in self-directed learning. For example, P91 said, ``I had to look up what the ord() function meant to help understand the code better.'' 


Others suggested that becoming skilled at prompting through use of the tool could lead to over-reliance on code generation and hinder their ability to write the code themselves.
\begin{quote}
    ``It is beneficial to my learning when I'm absolutely stuck on something, however, it does lead to me cutting corners and not putting as much thought into problems as I otherwise would.'' (P99)
\end{quote}


%
%

\section{Discussion}
\label{sec:discussion}

\subsection{Interaction Patterns}

From the data presented above, we observed three novel interaction patterns with Prompt Problems via \textsc{Promptly} that we believe illustrate their potential as an important pedagogical device.

\subsubsection{Play}
The data and examples listed in \ref{subsec:quant-playing} indicate that many students played with the tool because they continued to submit new variations on their prompts after successfully solving the problem. Additionally, while not part of our qualitative coding, we did find that students self-reported having fun and enjoyment using the tool. They wrote things like "The exercises were pretty fun and I enjoyed doing it" and "I really enjoyed this exercise and how it integrated generative artificial intelligence (AI) with Python programming language" and "This was fun [smile]". Because most types of introductory programming assignments are in high-level languages and written in complete syntax, they do not easily lend themselves to play. A notable exception to this is for ``toy'' interfaces, such as Scratch \cite{maloney2008programming}. Some novel interfaces for learning AI involve play, such as making and remixing music \cite{frid2020music}, but this does not teach programming concepts directly. Other types of novel interfaces that combine play with learning how to program focus on teaching abstract concepts like loops or conditionals \cite{lode2013machineers,melcer2018bots}. While not the same kind of play as a game, learning programming through Prompt Problems via tools such as \textsc{Promptly} combines learning high-level coding languages with the playful interaction style of talking to an intelligent agent in a novel way. However, the extent to which students find Prompt Problems playful needs further validation.

\subsubsection{Iteration}
Most students did as we intended: they prompted, thought about the results, and iterated. But we were surprised by the level of engagement indicated by students who continued to iterate longer than expected as seen in Section \ref{subsec:quant-long-tail}. The qualitative results in Section \ref{subsec:qual-prompting} further illuminate this interaction pattern of iteration on Prompt Problems. Even though some students found it tedious, others found it similar to writing code (see above, P227). Teaching programming through iteration with the so-called ``editing-running-debugging loop'' has been a well-known paradigm for decades. Doing this process over and over again through multiple programs is the gold standard of programming education \cite{allen2019many, denny2011codewrite}. However, as Guibert et al. pointed out in 2004, the lack of interactivity in this ``editing-running-debugging loop'' is a major aggravating factor \cite{guibert2004example}. Solutions to this usually involve engaging programming in some other way, such as new toy languages that abstract away the difficulties of syntax \cite{nishimura2014takt,weidmann2022bridging}. Furthermore, modern LLMs interrupt this cycle when students can copy and paste the program description into a prompt and have it solve the problem immediately. Our data suggests that Prompt Problems take a major step toward solving these issues by retaining the 
longstanding educational loop while adapting it for modern programming powered by LLMs.

\subsubsection{Reflection}
The qualitative results described above in Section \ref{subsec:qual-learning} indicate that students reflected on their prompts and the results, learned new concepts, and verbalized metacognition. Novice programmers consistently struggle with learning syntax and computational thinking while also learning and utilizing metacogntiive skills \cite{prather2018metacognitive, prather2020we,loksa2022metacognition}. Traditional solutions are to scaffold students directly \cite{loksa2016programming} or indirectly \cite{prather2019first}. One interesting recent example from Pechorina et al. involved a tool that explicitly guided novice programmers through the six stages of programming problem solving \cite{pechorina2023metacodenition}. While they showed interesting and positive results, manually moving through these stages for dozens of programming assignments throughout a course is highly constraining. Furthermore, although it teaches metacognition well, it does not address concerns about students using LLMs to solve their programming problems. Prompt Problems addresses these issues by affording students the ability to reflect on the code generated from their explanation, their progress toward a solution, and the relationship between their explanation and the generated code, all as a built-in part of the process. It is scalable and can augment different topics found in typical introductory programming courses.

\subsection{Designing Prompt Problems}
\label{sec:designingpromptproblems}
\begin{figure}
\centering
  \includegraphics[width=.7\linewidth]{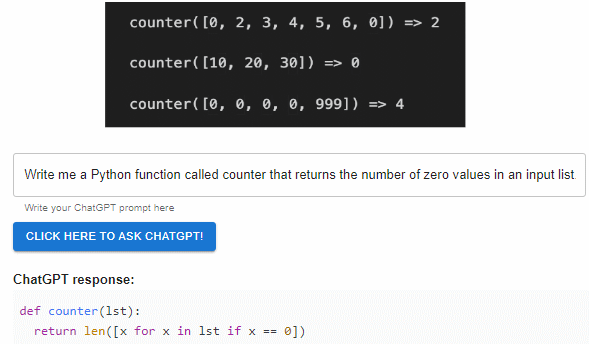}
  \caption{A problem in \textsc{Promptly} represented as a set of input-output pairs, where the solution requires generating a function (called `counter') that returns a count of the number of times zero occurs in a list. One possible prompt, and the resulting code that correctly solves the problem, is shown. }
  \label{fig:function_example}
\end{figure}

In our design of \textsc{Promptly}, we use graphical representations of problems so that students are tasked with crafting the text for the prompt themselves, and so that content from the problem representation cannot be easily copied and pasted into the prompt.  Figure \ref{fig:function_example} illustrates an example of an image used to represent a problem.  We found there were a number of considerations to keep in mind when designing Prompt Problems.

First, it is not straightforward to design complex problems without using any textual cues.  In some cases we included a graphical depiction alongside the test cases to illustrate the problem, for example the representation of judges holding score cards (see Figure \ref{fig:ex3_question}) to illustrate that the minimum and maximum values should be ignored as is common in judging athletic competitions.  Naturally, this assumes some domain knowledge that not all students may have, and we observed that this problem was the most difficult in our pilot study.  

Second, designing image-only questions has accessibility implications.  For example, students who have visual impairments typically rely on text-based descriptions of media.  However, if a text-description is provided for a Prompt Problem image, it could either describe the problem in sufficient detail to be passed directly to an LLM, bypassing student involvement in prompt creation, or it may add more complexity to the inductive reasoning needed to understand the problem illustrated by the visualization.  This could be addressed by providing an audio recording of a clear description of all the details of the image.

Third, English-native speaking students may produce more nuanced prompts and have greater success in improving them, potentially disadvantaging those with limited English skills in manipulating LLMs, even if they understand the problem well.  Therefore, if incorporating prompt generation activities into summative assessments, instructors should take this potential bias into account. 

Finally, when solving Prompt Problems, the underlying LLM might generate advanced code which is unsuitable for the course and which could confuse or demotivate learners.  Indeed, we observed this in our qualitative analysis of students' perceptions, and some student commented that the generated code was ``too advanced'' for them.  This is an issue that could be mitigated through tool design, for example by adding filters to allow or restrict the use of certain programming constructs that could be configured by the instructor. 

Prompt creation is a new kind of task that, as a community, we do not have much experience with.  Although Prompt Problems may be a useful vehicle for developing prompting skills in students, instructors may initially find it difficult to predict how hard it will be for students to solve a variety of them.  Further thought is needed about when to introduce these tasks, and other instruction related to prompting LLMs (such as biases in AI), into the curriculum.  


\subsection{Prompt Problems and the Future of Computing Education}
Prompt problems are a novel type of LLM-enabled problem, for students and instructors. Across the computing curriculum, students are typically exposed to both `bottom-up' and `top-down' approaches to learning. For example, a programming course might follow a bottom-up approach and begin teaching low-level syntax and semantics of a particular language, and then build up to more complex concepts such as algorithms and design patterns. The Prompt Problems that we explore in this work combine aspects of both approaches. On the one hand, they can be considered bottom-up, given that students start with input-output pairs and have to infer a problem description.  On the other, they are then expected to abstract the meaning of the input-output pairs into English sentences (prompts) rather than to code. This more top-down aspect requires students to understand the problem before they can generate prompts for the LLM to produce the correct code. 

The goal of Prompt Problems is to help students learn how to create effective prompts for LLMs.  With a tool like \textsc{Promptly}, which automatically verifies solutions for each problem through the use of pre-defined unit tests, students are able to focus on constructing the prompt.  However, in practice, when LLMs are used to generate code it is the programmer's responsibility to check that the generated code is correct.  Thus, testing is a key programming skill for students to develop~\cite{janzen2006tdd}.  Future tools that focus on prompt generation may benefit from the integration of user-generated tests, such as requiring students to create one or more of their own tests before beginning to construct prompts.  Not only can this help to develop testing skills, but generating test cases prior to writing code is a well-known strategy for improving problem understanding \cite{pechorina2023metacodenition, denny2019closer}.

The inevitable increase in use of LLMs in the classroom, for which Prompt Problems are just one example task, will likely lead to a shifting emphasis in certain traditional skills. For example, while the importance of learning to read code has been emphasized in computing education for decades~\cite{lopez2008relationships}, the advent of AI code generators suggests this skill will becoming increasingly important.  Students are likely to spend relatively more programming time on reading AI-suggested code~\cite{becker2023programming}. Traditionally, introductory programming courses have heavily focused on code writing tasks, and less on reading code produced by others. Prompt Problems could help in this changing landscape as they expose students to code generated by LLMs in a way where critically reading and evaluating the code can be a useful strategy for improving subsequent prompts. 

In traditional programming exercises, students are usually quite familiar with the process of following the provided specifications for a problem, but are less familiar with the process of carefully specifying a problem statement.  Indeed, in our study students found it challenging to construct prompts with appropriate specificity and to translate between their understanding of the problem and suitable prompts. Common strategies we observed included trying variations on ideas by changing a few words, as opposed to continually adding more words. Many students, particularly those with long submission streams, also made attempts to reduce the total number of words in their prompts.

\subsection{Limitations}
\label{sec:limitations}
There are several limitations to our studies. The first one is that we do not provide students with any explicit feedback about how to create effective prompts.  Instead, we rely on students intuitions about how to craft good prompts, and the prompts are evaluated purely on whether or not they lead to a code solution that passes the test cases.  However, a prompt could be overly verbose, or include irrelevant instructions that generate unnecessary code constructs, and still generate code that produces the expected outputs.  See, for example, the wide variety of successful prompts (e.g. P175 and P298) in Section \ref{sec:results:quant}.  Future work should address this issue, by analyzing prior successful prompts to a problem, and provide feedback to students about not just prompt correctness -- but also other measures.  One important measure is prompt robustness.  LLMs are non-deterministic, and thus generate different outputs for the same input.  Ideally, a prompt should be robust to such variations, and generate correct code consistently.  This could be incorporated into the tool directly, by generating multiple LLM outputs rather than just one and evaluating a prompt as robust only if all of these outputs pass the tests.


Second, we do not know if the students being sampled have equal background, knowledge, and experience with prompting. Even though we did not collect this data, anecdotally it appears they do not have equal background in prompting. Some students self-reported that they won't ever use generative AI, while some said they use it frequently. Students said things like "actually, I never use AI or ChatGPT. i never use ai before, even any optional question" (P131) and "I find it useful, but when I use it, I gaslight myself into believing that I already know what it says" (P141) and "The tool took a bit of getting used to at the very beginning however as I used it more, it became easier to get it to generate what I wanted" (317). Future work should explore how prior exposure to prompting changes outcomes and how this could be mitigated by classroom instruction to bridge accessibility and inclusion gaps.

Third, even though the present work is debuting a new class of programming assignment that could have a positive impact on the way computing is taught, it requires some specialized setup. Currently, to utilize Prompt Problems requires a tool that can accept student prompts, send them to a generative AI, return the feedback, and check for correctness. This may lower uptake of Prompt Problems in classrooms. However, other types of programming problems, such as Parsons Problems\cite{ericson2022parsons}, originally needed specialized implementation to utilize. Now, there are several tools freely available for Parsons Problems. We hope to see Prompt Problems follow a similar trajectory.

Fourth, our interface restricted users to single prompts without the ability to rely on previous prompts. However, this is not the way most people will encounter generative AI. Interfaces such as ChatGPT allow for conversations where the model can utilize the context of the thread to make better suggestions. We specifically chose a single prompt model, rather than a chain prompt model, because we wanted users to sharpen their prompting skills without relying on the model to help them via context. One goal of Prompt Problems is to get better at crafting excellent prompts the first time. In doing so, we were able to examine repeated prompts over time, illuminating the ways students tend to modify their prompts.

Finally, all the studies (exploratory study, pilot study, large scale study) were conducted at a single university and so therefore may lack generalizability. We attempted to mitigate this by collecting data in stages and then attempting to collect data with a large enough \emph{n}. This work should be replicated in other contexts and other countries. We also based many of our tool design decisions on the exploratory and the pilot studies, which were non-assessed optional assignments. This may have biased the results of these exploratory and pilot studies towards students who were more engaged or who like to try new things. However, we believe that the data in Section~\ref{sec:large-scale-study} supports our design decisions. Lastly, as discussed above, there are other ways to design Prompt Problems. Our design may have unintentionally biased our results where other implementations might not produce useful or helpful experiences for students. Future work should investigate these other design opportunities.

\section{Conclusion}
In this paper, we introduced a novel type of programming problem, called Prompt Problems, delivered via our tool, \textsc{Promptly}. Presented with the problem visually, students must decompose the problem into plain English, submit it to a generative AI model, read the code it returns and the test case data from the tool, modify their prompt based on this feedback, and continue iteratively toward a solution. The data we presented supports the idea that Prompt Problems are a useful way to teach programming concepts and encourage metacognitive programming skills while embracing the impact of LLMs on computing education.


\bibliographystyle{ACM-Reference-Format}
\bibliography{main}

\end{document}